%% file: main.tex
\documentclass[10pt,conference]{IEEEtran} 
\IEEEoverridecommandlockouts

\usepackage{url} 
\usepackage{framed}
\usepackage{mdframed}
\usepackage{listings}
\usepackage{multicol}
\usepackage{amssymb}
\usepackage{lipsum}
\usepackage{adjustbox}
\usepackage[font=bf,skip=\baselineskip]{caption}
\usepackage{tabularx}
\usepackage{seqsplit}
\usepackage{multirow}
\usepackage{booktabs}
\usepackage{graphicx}
\usepackage{hyperref}
\usepackage{subcaption}
\usepackage{makecell}
\usepackage[table]{xcolor}
\usepackage[ruled,vlined,linesnumbered]{algorithm2e}
\usepackage{amsmath}
\usepackage{tikz}
\usepackage{tcolorbox}
\usepackage{threeparttable}

\usepackage{breakurl} 
\usepackage{boldline}
\usepackage[ruled,vlined]{algorithm2e}

\usepackage{pifont}
\usepackage{balance}
\let\labelindent\relax
\usepackage{enumitem}
\usepackage[htt]{hyphenat}
\usepackage{array}

\definecolor{customblue}{HTML}{006ca6}
\definecolor{customgreen}{HTML}{009264}
\definecolor{custombrown}{HTML}{ff3d00}
\AtEndPreamble{
\usepackage{hyperref}

    \hypersetup{
      colorlinks = true,
      linkcolor = customgreen,
      anchorcolor = purple,
      citecolor = customgreen,
      filecolor = purple,
      urlcolor = customblue
    }
}

\newcommand{\find}[1]{
\begin{tcolorbox}[leftrule=0.5mm,toprule=0mm,bottomrule=0mm,left=0.7pt,right=0.7pt,top=0.2pt,bottom=0.2pt]
\em #1
\end{tcolorbox}
}

\newcommand{\ea}{et~al.}

\begin{document}

\title{Voices from the Frontier: A Comprehensive Analysis of the OpenAI Developer Forum}

\author{
\IEEEauthorblockN{Xinyi Hou\IEEEauthorrefmark{1}, Yanjie Zhao\IEEEauthorrefmark{1}, and Haoyu Wang\IEEEauthorrefmark{2}}
\IEEEauthorblockA{
Huazhong University of Science and Technology, Wuhan, China\\
xinyihou@hust.edu.cn, yanjie\_zhao@hust.edu.cn, haoyuwang@hust.edu.cn}
\thanks{\IEEEauthorrefmark{1}Xinyi Hou and Yanjie Zhao contributed equally to this work.}
\thanks{\IEEEauthorrefmark{2}Haoyu Wang is the corresponding author (haoyuwang@hust.edu.cn).}
}

\maketitle

\begin{abstract}

OpenAI's advanced large language models (LLMs) have revolutionized natural language processing and enabled developers to create innovative applications. As adoption grows, understanding the experiences and challenges of developers working with these technologies is crucial. This paper presents a comprehensive analysis of the OpenAI Developer Forum, focusing on (1) popularity trends and user engagement patterns, and (2) a taxonomy of challenges and concerns faced by developers. 
We first employ a quantitative analysis of the metadata from 29,576 forum topics, investigating temporal trends in topic creation, the popularity of topics across different categories, and user contributions at various trust levels. We then qualitatively analyze content from 9,301 recently active topics on developer concerns. From a sample of 886 topics, we construct a taxonomy of concerns in the OpenAI Developer Forum. Our findings uncover critical concerns raised by developers in creating AI-powered applications and offer targeted recommendations to address them. This work not only advances AI-assisted software engineering but also empowers developer communities to shape the responsible evolution and integration of AI technology in society.

\end{abstract}

\input{Sections/1.introduction}

\input{Sections/2.background}
\input{Sections/3.methodology}

\input{Sections/4.rq1}

\input{Sections/4.rq2}

\input{Sections/5.discussion}

\input{Sections/6.limitation}

\input{Sections/8.conclusion}


\bibliographystyle{IEEEtranS}
\bibliography{main}

\end{document}

%% file: Sections/1.introduction.tex
\section{Introduction}
\label{sec:introduction}

\begin{figure*}
    \centering
    \includegraphics[width=1\linewidth]{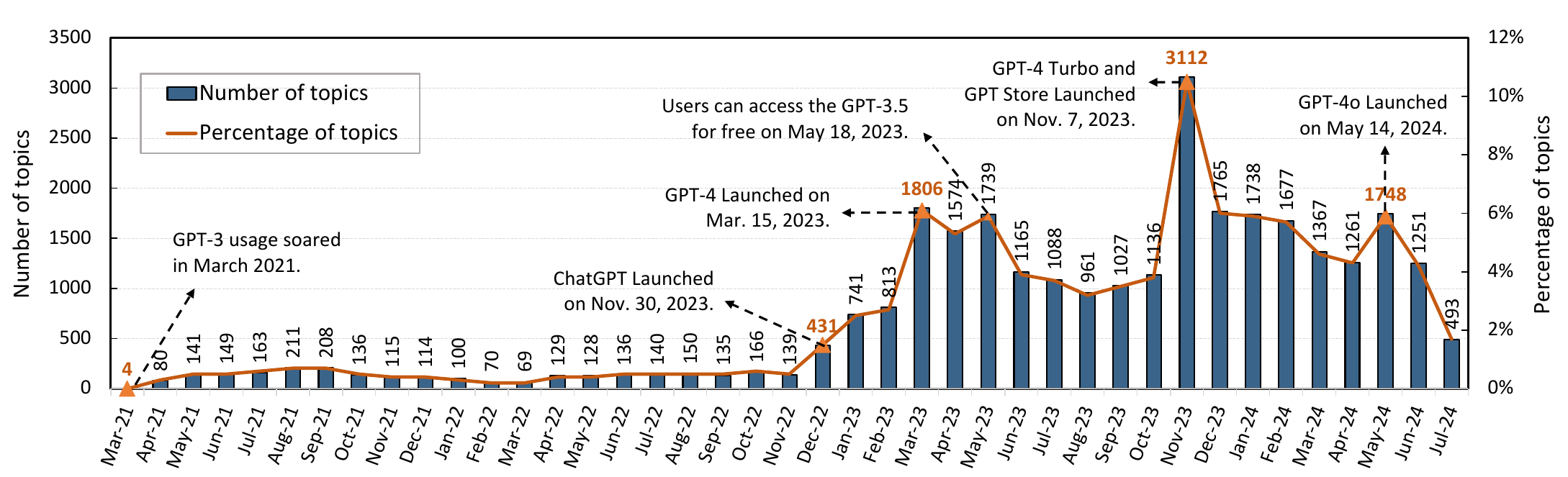}
    \caption{Trend in topic counts on OpenAI Developer Forum over time (Up to July 16, 2024).}
    \label{fig:topic_time}
\end{figure*}

OpenAI~\cite{openai}, a leading AI research organization, has made significant contributions to the field of natural language processing with the development of large language models (LLMs) like GPT-4~\cite{achiam2023gpt}. These models have demonstrated remarkable capabilities in language understanding, generation, and task-solving. The release of ChatGPT~\cite{chatgpt}, a conversational AI system, has further sparked interest and engagement from the developer community. To support this growing interest, OpenAI provides a rich suite of tools and resources to streamline AI integration~\cite{developerplatform}, including the \texttt{Assistant API}~\cite{assistants}, \texttt{.NET SDK}~\cite{sdk}, \texttt{Vector databases}~\cite{vectordatabase}, etc. Additionally, the GPT Store~\cite{gptstore} allows developers to create and share custom GPTs, fostering a vibrant ecosystem of innovation. As developers increasingly leverage OpenAI's models to build applications across various domains, they encounter a range of challenges and concerns specific to working with these AI systems. These challenges \textbf{span technical issues related to model integration, prompt engineering, and output quality, as well as broader ethical considerations about bias, fairness, and responsible deployment}. Understanding and addressing these challenges is crucial for the software engineering (SE) community to ensure the development of robust, reliable AI-powered applications.

The OpenAI Developer Forum~\cite{openai2024forum} serves as a primary platform for developers to discuss their experiences, seek guidance, and share best practices related to working with OpenAI's technologies. \autoref{fig:topic_time} illustrates the trend in topic counts on the forum over time, with notable spikes coinciding with key events such as the launch of ChatGPT, GPT-4, and GPT Store. This trend suggests that the forum acts as a barometer for developer interest and engagement with OpenAI's technologies, making it a valuable resource for understanding the real-world challenges developers face and identifying areas where additional research, tools, and support are needed. In this paper, we are the first to present a comprehensive analysis of the OpenAI Developer Forum, focusing on two main research questions (RQs):

\noindent\textbf{RQ1: Popularity Trends.} RQ1 aims to conduct a comprehensive analysis of the OpenAI Developer Forum's overall activity. We seek to provide a foundational understanding of the forum's dynamics, highlighting popular topics and user engagement trends, which is essential for identifying key areas of interest within the developer community.

\noindent\textbf{RQ2: Taxonomy of Concerns.} RQ2 focuses on classifying and analyzing the specific concerns raised by developers in the OpenAI Developer Forum. Identifying these concerns will help in understanding the common obstacles and issues developers encounter, offering valuable insights that can improve best practices and support the SE community.

To address RQ1, we quantitatively analyze forum metadata, examining topic creation times, category distribution, post scores, and active user numbers, providing insights into popularity trends and user engagement. For RQ2, we perform a qualitative analysis of topics in the forum to develop a taxonomy that categorizes the practical challenges and broader concerns of developers. Our contributions include:

\begin{enumerate}
    \item We collected 29,576 topics from the OpenAI Developer Forum and provided an in-depth analysis of the forum's popularity trends, highlighting key areas of interest and user engagement patterns.
    \item We filtered 9,301 recently active topics related to developers' challenges and concerns on the OpenAI Developer Forum. We then constructed a comprehensive taxonomy of these discussions based on a sample of 886 topics.
    \item We are the first to conduct a large-scale OpenAI Developer Forum analysis, providing insights that guide future research, tool development, and best practices in AI-assisted software engineering.
\end{enumerate}

%% file: Sections/2.background.tex
\section{Background and Related Work}
\label{sec:backgroud}

\input{Tables/category_description}

\subsection{OpenAI Developer Forum}

The OpenAI Developer Forum~\cite{openai2024forum} is a pivotal platform where LLM developers, researchers, and enthusiasts converge to discuss, troubleshoot, and share insights on various aspects of LLM technology. As one of the most active online communities dedicated to LLMs, the forum serves as a microcosm of the broader trends, challenges, and concerns faced in the field of LLM for SE~\cite{hou2023large}. 
The forum is divided into main categories, as shown in \autoref{tab:topic_categories}.
These include: \texttt{Announcements} for updates from OpenAI; \texttt{API} for discussing bugs, feedback, and deprecations of OpenAI's APIs; \texttt{ChatGPT} for support and feature requests of ChatGPT; \texttt{Community} for general interactions; \texttt{Forum feedback} for platform improvement suggestions; \texttt{GPT builders} for topics on the GPT Store and GPT development; \texttt{Documentation} for discussions on OpenAI's resources; and \texttt{Prompting} for optimizing AI interactions.
The OpenAI Developer Forum features a diverse community, from novices to leading researchers. This diversity promotes knowledge exchange, allowing beginners to learn from experts and experts to stay updated on practical challenges. The forum's collaborative nature encourages knowledge sharing, creating a dynamic knowledge base. For researchers, \textbf{it offers insights into real-world LLM usage, helping identify common issues and technology gaps}. This grounds theoretical research in practical needs, aligning academic progress with developer community requirements.

\subsection{Related Work on Forum Analysis}

Online forums in the SE community, like Stack Overflow, contain a wealth of information about problems encountered during software development, solutions adopted, developer opinions, and experiences. 
Numerous studies have analyzed online forum data in SE communities to understand developers' challenges. Rosen~\ea~\cite{rosen2016mobile} investigated mobile developers' questions on Stack Overflow, providing insights for research and tool improvements. Abdellatif~\ea~\cite{abdellatif2020challenges} studied chatbot development challenges through Stack Overflow posts, while Wan~\ea~\cite{wan2019programmers} examined blockchain platform discussions across Stack Exchange communities. Zhang~\ea~\cite{zhang2019empirical} and Han~\ea~\cite{han2020programmers} investigated challenges in deep learning applications and frameworks. Venkatesh~\ea~\cite{venkatesh2016client} analyzed client developers' concerns when using web APIs, and Ahmed~\ea~\cite{ahmed2018concurrency} explored concurrency-related questions on Stack Overflow. 

These studies guide research directions, tool development, and best practices in various SE domains.
Unfortunately, \textbf{no research has yet analyzed the OpenAI Developer Forum}. This gap motivates our study, which examines the experiences and challenges of developers working with cutting-edge AI technologies. Our work provides insights to help SE researchers identify key areas needing further investigation and support for effective AI adoption in software development.

%% file: Tables/category_description.tex
\begin{table*}[h!]
\centering
\caption{Topic categories on OpenAI Developer Forum.}
\resizebox{0.98\linewidth}{!}{
\begin{tabular}{lll}
\toprule[1.2pt]
\textbf{Topic Category} & \textbf{Description} & \textbf{Subcategory} \\ \midrule
Announcements    & Official updates related to OpenAI, the API, ChatGPT, and more. & / \\ 
API             & Questions, feedback, and best practices around building with OpenAI’s API. & Bugs, Feedback, Deprecations \\ 
ChatGPT        & Questions or discussions about ChatGPT. &  Bugs, Feature requests, Use cases, Support\\ 
Community      & A place to connect with the OpenAI Developer community. & / \\ 
Forum feedback   & Feedback on how to make this developer forum better for users. & / \\ 
GPT builders     & Create tailored versions of ChatGPT for specific tasks and share them. & Plugins \& Actions builders, Plugin store\\ 
Documentation    & Share feedback on documentation and tutorials about OpenAI. & / \\ 
Prompting       & Learn more about prompting by sharing best practices and more. & / \\ 
\bottomrule[1.2pt]
\end{tabular}
}
\label{tab:topic_categories}
\end{table*}

%% file: Sections/3.methodology.tex
\section{Methodology}
\label{sec:methodology}

\begin{figure}[h!]
    \centering
    \includegraphics[width=0.95\linewidth]{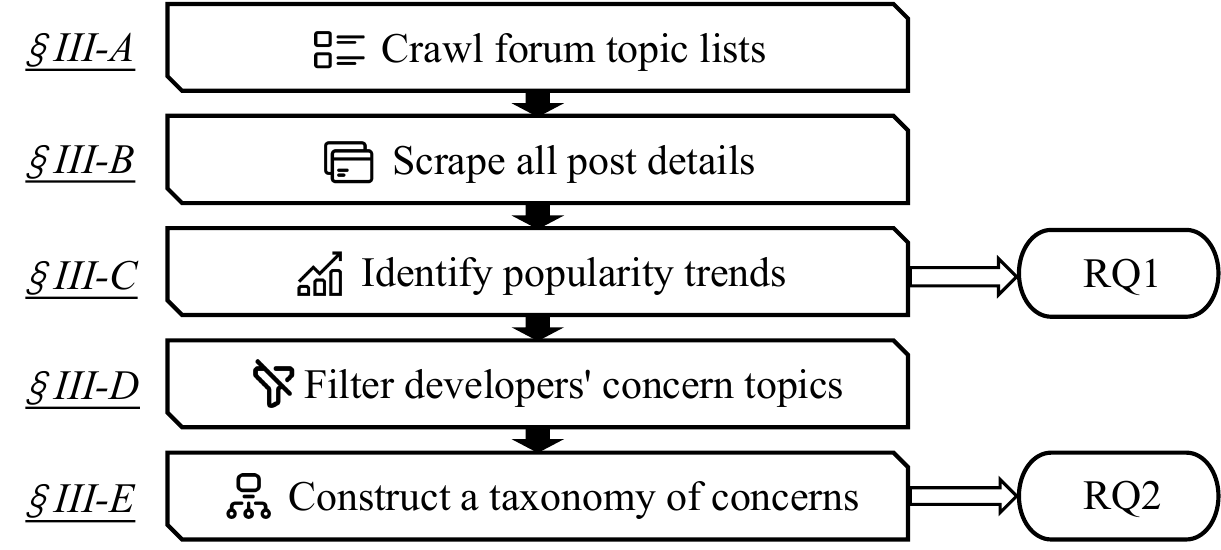}
    \caption{Overview of methodology.}
    \label{fig:methodology}
\end{figure}

This section outlines the systematic methodology used to analyze the OpenAI Developer Forum, which consists of five primary steps, as illustrated in \autoref{fig:methodology}.

\subsection{Crawl Forum Topic Lists}
To initiate our study, we developed a custom web crawler to retrieve comprehensive topic lists from the OpenAI Developer Forum. All topics must be assigned to a specific category upon creation, allowing us to systematically traverse the 17 categories listed in \autoref{tab:topic_categories} (eight main categories and nine subcategories) to compile the complete topic list of the OpenAI Developer Forum. We identified the backend API endpoints
for each category and paginated through them to collect basic information about each topic, including 33 attributes such as \textit{topic\_id}, \textit{title}, \textit{topic\_slug}, \textit{posts\_count}, \textit{created\_at}, and \textit{tags}. To verify the completeness of our retrieved topic list, we also crawled all topics from the homepage under the ``Latest'', ``Top'', and ``Hot'' sections, which are not restricted by topic categories. The final topic count obtained from these four different crawling methods was consistent, confirming the comprehensiveness of our topic list. Through this systematic collection process, we retrieved a total of 29,576 topics by July 16, 2024, which not only facilitated the subsequent scraping of posts but also provided data to support the analysis of popularity trends and developer concerns.

\subsection{Scrape All Post Details}

Based on the retrieved \textit{topic\_id}, \textit{topic\_slug}, and \textit{posts\_count}, we constructed precise URLs for all topic pages.
We then used automated scraping tools to extract complete post information for each topic, including 52 attributes such as \textit{post\_id}, \textit{username}, \textit{cooked} (the post's actual content), \textit{post\_type}, \textit{incoming\_link\_count}, \textit{readers\_count}, \textit{trust\_level}, and \textit{score}.

\subsection{Identify Popularity Trends}

To identify trends in topic popularity and user engagement, we conducted a time series analysis to investigate the growth and decline of interest in specific topics over time. Additionally, we employed statistical methods to examine the relationships between various factors (e.g., topic category, author reputation, and post sentiment) and the level of engagement received by individual posts. 
These analyses provided valuable insights into the dynamics of popularity within the OpenAI Developer Forum.

\subsection{Filter Developers' Concern Topics}

To understand the current concerns of developers, we filtered the collected 29,576 topics through a four-step process. First, to ensure the topicality of the discussions, we retained 12,440 topics that were last updated in 2024 based on the \textit{last\_posted\_at} attribute. Second, to ensure these topics are currently accessible to the public, we filtered down to 10,556 topics using the \textit{closed} attribute. Given that the OpenAI Developer Forum is an active community with developers and OpenAI staff actively participating, we then focused on 9,417 topics that had not yet received an accepted answer, as indicated by the \textit{has\_accepted\_answer} attribute. As shown in \autoref{tab:topic_categories}, the forum comprises various categories, but \texttt{Announcements}, \texttt{Community}, \texttt{Forum feedback}, and \texttt{Documentation} mainly address product launches and community building, not specific developer concerns. 
Therefore, we focused our analysis on the remaining four categories: \texttt{API}, \texttt{ChatGPT}, \texttt{GPT builders}, and \texttt{Prompting}. After this final filtering step, we were left with 9,301 topics, distributed as follows: 5,865 topics in \texttt{API}, 1,724 topics in \texttt{ChatGPT}, 1,113 topics in \texttt{GPT builders}, and 599 topics in \texttt{Prompting}.

\subsection{Construct a Taxonomy of Concerns}

We constructed a taxonomy of concerns based on the OpenAI Developer Forum's official categories: \texttt{API/ChatGPT}, \texttt{GPT builders}, and \texttt{Prompting}. Given the similarity in topics between \texttt{API} and \texttt{ChatGPT} in the forum and the frequent comparisons made by developers between API and ChatGPT, we combined these two categories for analysis. According to Zhang~\ea~\cite{zhang2019empirical}, to ensure a 95\% confidence level and a 5\% confidence interval, we sampled a total of 886 topics from the three categories (\texttt{API/ChatGPT} with 366 topics, \texttt{GPT Builders} with 286 topics, and \texttt{Prompting} with 234 topics) for constructing the taxonomy. The taxonomy construction process is described below.

\noindent\textbf{Manual preliminary construction}.
Utilizing an open coding procedure~\cite{seaman1999qualitative}, we inductively developed categories and subcategories for the taxonomy by analyzing topics from the OpenAI Developer Forum.
Two researchers (referred to as inspectors), who both possess extensive experience in LLM development and API usage, collaborated on the preliminary construction. We randomly sampled 30\%~\cite{chen2020comprehensive} of the 886 forum topics for this initial phase.

The inspectors thoroughly reviewed all sampled topics, taking into account the title, tags, body, replies, and any URLs referenced by forum participants. Topics unrelated to developer concerns, such as those promoting GPTs, were not retained. For the rest of the topics, the inspectors assigned brief phrases as initial codes to represent the underlying concerns related to API/ChatGPT usage, GPT builders, and prompting.
The inspectors then grouped similar codes into categories, forming a hierarchical taxonomy that addresses concerns specific to OpenAI's products and services.
This grouping process was iterative, with inspectors continuously refining the taxonomy by moving between categories and forum topics. Topics related to multiple concerns were assigned to all relevant categories. All disagreements were resolved through discussion with an experienced arbitrator knowledgeable of OpenAI's ecosystem.

\noindent\textbf{Reliability analysis and extended construction}.
Based on the coding schema from the preliminary construction, the two inspectors independently coded the remaining 70\%~\cite{chen2020comprehensive} of forum topics for reliability analysis. 
Each topic was assigned to the identified leaf categories in the taxonomy or discarded for being unrelated to developer concerns. 
Topics that could not be classified within the existing taxonomy were placed into a newly created category called \texttt{pending}.
The independent labeling process yielded an inter-rater agreement of 0.835, as measured by Cohen's Kappa~(\(\kappa\)). This high level of agreement signifies almost perfect concordance, highlighting the reliability of our coding schema.
Coding conflicts were discussed and resolved with the arbitrator.
For \texttt{pending} topics, the arbitrator assisted in identifying the underlying concerns and deciding if new categories were required.
Finally, six additional leaf categories were created, and all topics previously classified as \texttt{pending} were assigned to the taxonomy. The entire manual construction process took roughly 200 person-hours.

In summary, among the 886 sampled topics, 169 were unrelated to developer concerns, and 717 topics were covered in the final taxonomy. The resulting taxonomy consists of three root categories (i.e., \texttt{API/ChatGPT}, \texttt{GPT Builders}, and \texttt{Prompting}), 13 inner categories, and 50 leaf categories, as shown in \autoref{fig:taxonomy}.

%% file: Sections/4.rq1.tex
\section{RQ1: Popularity Trends}
\label{sec:rq1}

This section explores the popularity trends within the OpenAI Developer Forum. By analyzing various aspects of forum activity, we aim to uncover patterns in user engagement and topic discussions. We specifically examine the temporal trends in topic creation (\autoref{sec:overtime}), the popularity of topics across different categories (\autoref{sec:bycategory}), and the contributions of users at various trust levels (\autoref{sec:trustlevel}).

\subsection{Topic Trends Over Time}
\label{sec:overtime}

The OpenAI Developer Forum has become a hub for developers to discuss cutting-edge advancements in AI and share their insights on the latest developments from OpenAI. As seen in \autoref{fig:topic_time}, the forum's popularity and user engagement have grown significantly over time, \textbf{with spikes in activity often coinciding with major announcements and releases from OpenAI}. Key events such as the launch of GPT-4 in March 2023, and the introduction of the GPT Store in November 2023 have sparked widespread discussions among developers on the forum. These spikes in activity demonstrate the keen interest and excitement within the developer community regarding OpenAI's groundbreaking advancements in language models and AI technologies. 
The OpenAI Developer Forum has clearly become a focal point for developers who are passionate about staying up-to-date with the latest developments in AI. The concerns, ideas, and insights shared by this vibrant community of developers can provide valuable inspiration and guidance for the broader SE community.

\subsection{Topic Popularity by Category}
\label{sec:bycategory}
The OpenAI Developer Forum organizes topics into eight main categories, as shown in \autoref{tab:topic_categories}. Among these categories, \texttt{API}, \texttt{ChatGPT}, and \texttt{GPT builders} contain several subcategories. \autoref{tab:topic_count} presents the distribution of topic counts across these categories. The \texttt{Announcements} category has the fewest number of topics, primarily because only official announcements can be posted in this category. The \texttt{API} category far surpasses other categories in terms of both topic and post counts, indicating that \textbf{discussions related to APIs generate the highest level of interest and engagement on the OpenAI Developer Forum.}

\input{Tables/tpoic_count}

\autoref{tab:topic_count} includes the Max Score and Avg. Score for each category. Forum posts are assigned scores, likely reflecting their popularity or relevance. These scores influence topic rankings in the Hot and Top channels. We hypothesize that scores are based on various factors like the number of reads, replies, quotes, incoming links, positive feedback, timestamps, and the user's trust level or role. Notably, the highest Max Score is in the \texttt{Community} category for the topic ``ChatPDF.com - Chat with any PDF using the new ChatGPT API'', indicating strong community interest in practical applications of OpenAI's technologies. The \texttt{Announcements} category has the highest Avg. Score, likely due to the important of official updates and the extensive discussions they generate.

The analysis of topic popularity by category provides insights into the interests and priorities of the OpenAI Developer Forum community. It highlights \textbf{the significance of API-related discussions and showcases the community's enthusiasm for real-world applications of AI technologies}. Furthermore, the examination of the scoring system sheds light on the factors that contribute to the visibility and engagement of posts within the forum.

\subsection{User Contributions by Trust Level}
\label{sec:trustlevel}
To further understand the popularity trends on the OpenAI Developer Forum, it is essential to examine the distribution of user contributions across different trust levels. The forum employs a trust level system~\cite{trustlevel} to categorize users based on their engagement and contributions. The trust levels are defined as follows: Level 0 (\texttt{Newuser}), Level 1 (\texttt{Basic}), Level 2 (\texttt{Member}), Level 3 (\texttt{Regular}), and Level 4 (\texttt{Leader}). Higher trust levels represent a greater degree of participation and influence within the community.

\begin{figure}[h!]
    \centering
    \begin{subfigure}[b]{0.48\linewidth}
        \centering
        \includegraphics[width=\linewidth]{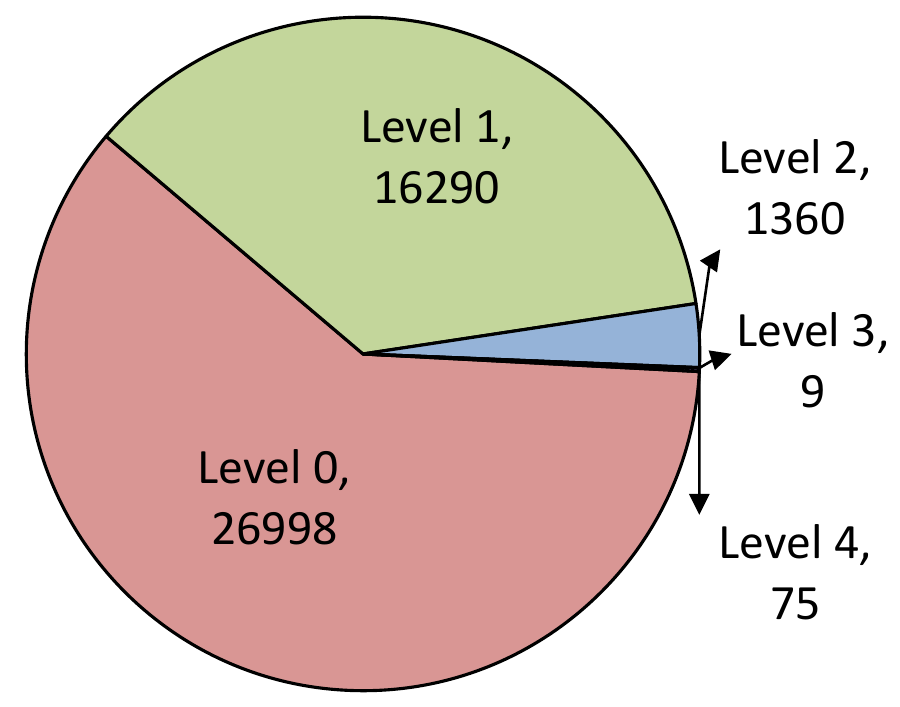}
        \caption{User distribution by trust level.}
        \label{fig:user_count}
    \end{subfigure}
    \hfill
    \begin{subfigure}[b]{0.48\linewidth}
        \centering
        \includegraphics[width=\linewidth]{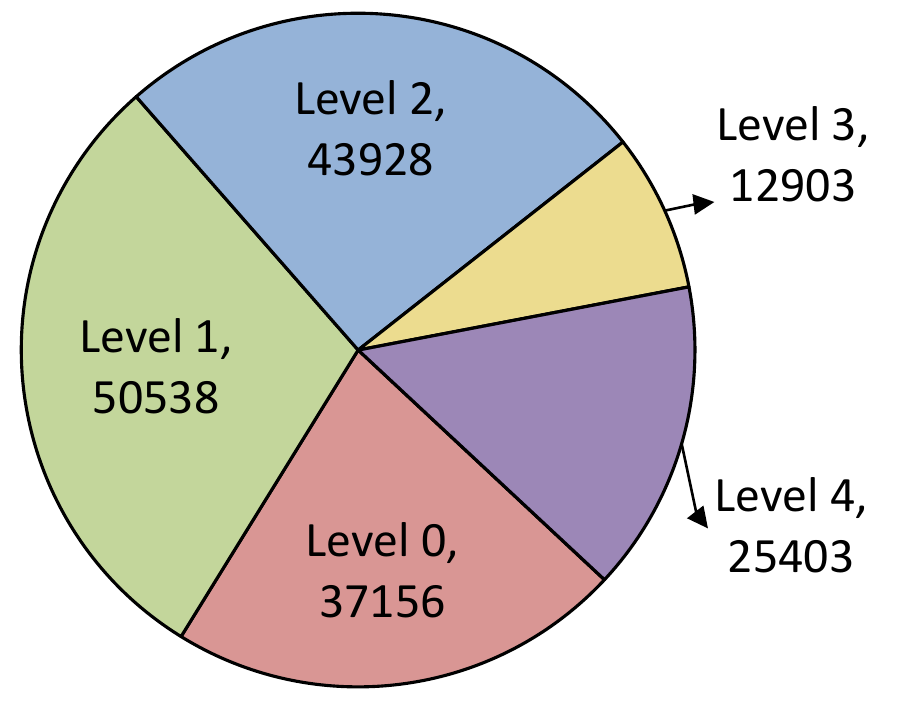}
        \caption{User post contributions by trust level.}
        \label{fig:user_post}
    \end{subfigure}
    \caption{User distribution and contributions by trust level.}
    \label{fig:user}
\end{figure}

\autoref{fig:user} illustrates the user distribution and post contributions by trust level. The majority of users (26,998) are in the \texttt{Newuser} category (trust level 0), but they account for a relatively small portion of the total post count (37,156). Interestingly, as trust levels increase, the number of users decreases significantly, but their individual contributions grow substantially. For instance, the 75 users in the \texttt{Leader} category (trust level 4) have authored an impressive 25,403 posts, highlighting \textbf{the pivotal role played by a small group of highly engaged and trusted users in driving discussions and generating valuable content}.

%% file: Tables/tpoic_count.tex
\begin{table}[h!]
\centering
\caption{Topic categories and their statistics.}
\resizebox{1\linewidth}{!}{
\begin{tabular}{lrrrr}
\toprule[1.2pt]
\textbf{Category} & \textbf{\# Topics} & \textbf{\# Posts} & \textbf{Max Score} & \textbf{Avg. Score} \\ \midrule
Announcements    & 35    & 1,077   & 289846.40  & 5068.01 \\ 
API              & 15,221 & 83,280  & 871119.40  & 890.53  \\ 
ChatGPT          & 5,433  & 30,195  & 1406922.80 & 922.79  \\ 
Community        & 3,499  & 23,361  & 4486547.40 & 987.42  \\ 
Forum feedback   & 65 & 348 & 22987.80   & 160.21  \\ 
GPT builders & 2,704 & 16,413  & 228394.40 & 354.28  \\ 
Documentation    & 336   & 1,594   & 157353.60  & 1073.32 \\ 
Prompting        & 2,283  & 13,660  & 4019560  & 833.03  \\ \midrule
\textbf{Total}   & 29,576 & 169,928 & 4486547.40 & 1286.20 \\ 
\bottomrule[1.2pt]
\end{tabular}}
\label{tab:topic_count}
\end{table}

%% file: Sections/4.rq2.tex
\section{RQ2: Taxonomy of Concerns}
\label{sec:rq2}

\begin{figure*}[h!]
    \centering
    \begin{subfigure}[b]{0.94\linewidth}
        \centering
        \includegraphics[width=\linewidth]{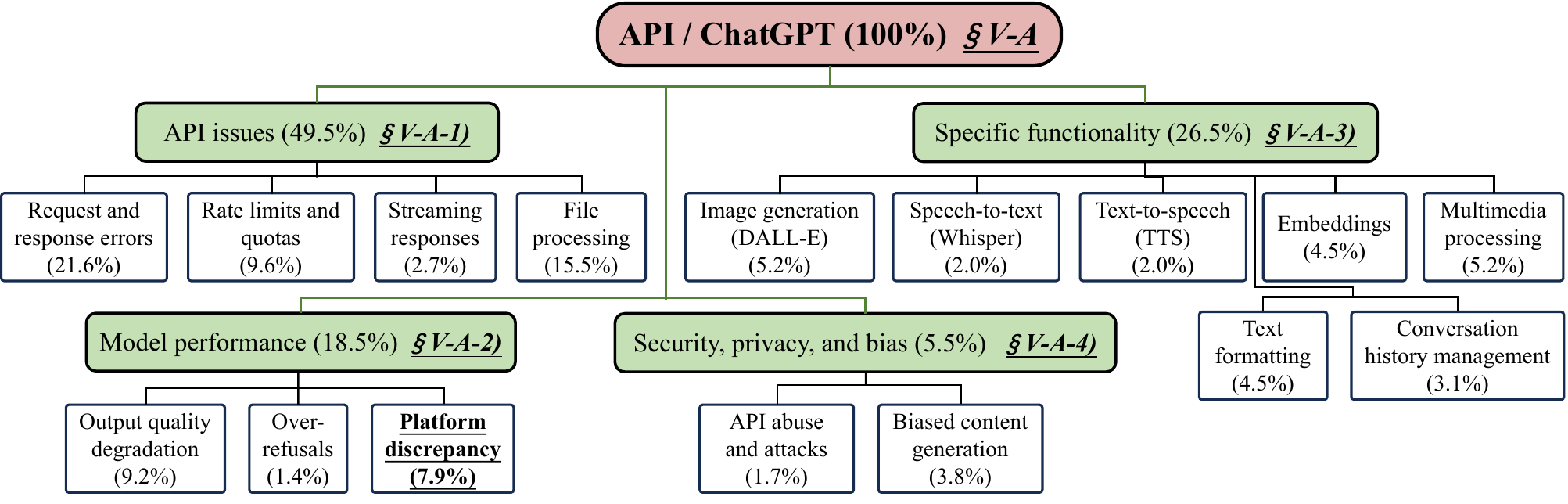}
        \caption{Taxonomy of concerns in \texttt{API} and \texttt{ChatGPT} topics.}
        \label{fig:api}
    \end{subfigure}
    
    \begin{subfigure}[b]{0.94\linewidth}
        \centering
        \includegraphics[width=\linewidth]{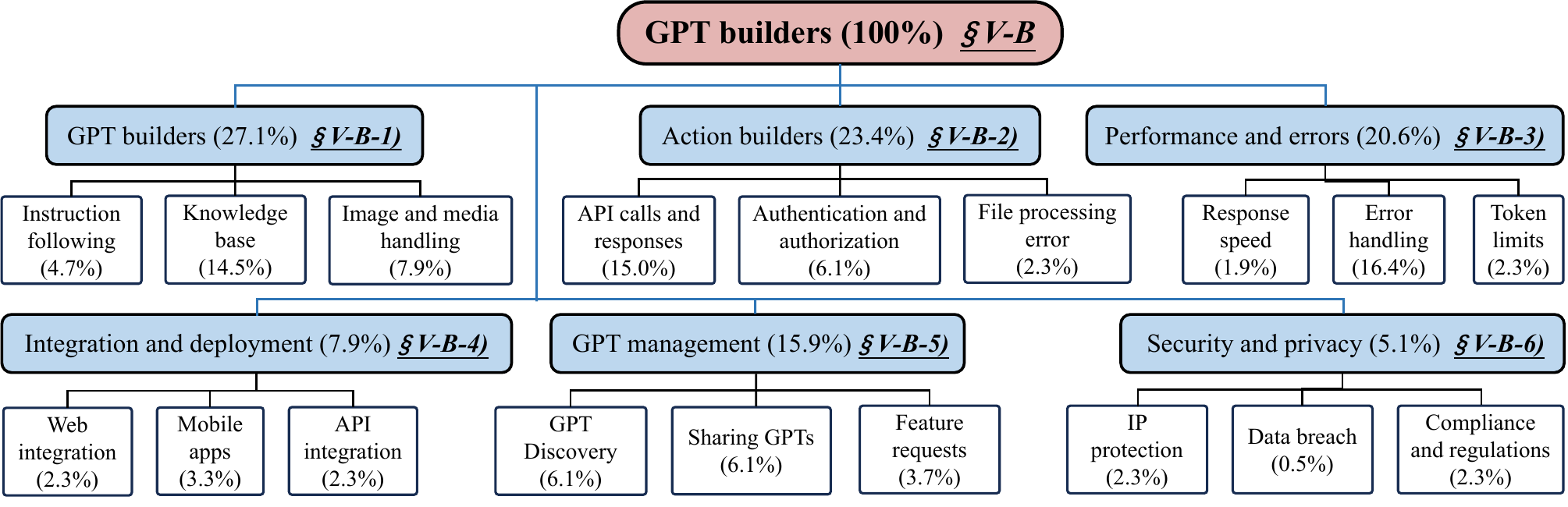}
        \caption{Taxonomy of concerns in \texttt{GPT builders} topics.}
        \label{fig:gpt_builders}
    \end{subfigure}
    
    \begin{subfigure}[b]{0.94\linewidth}
        \centering
        \includegraphics[width=\linewidth]{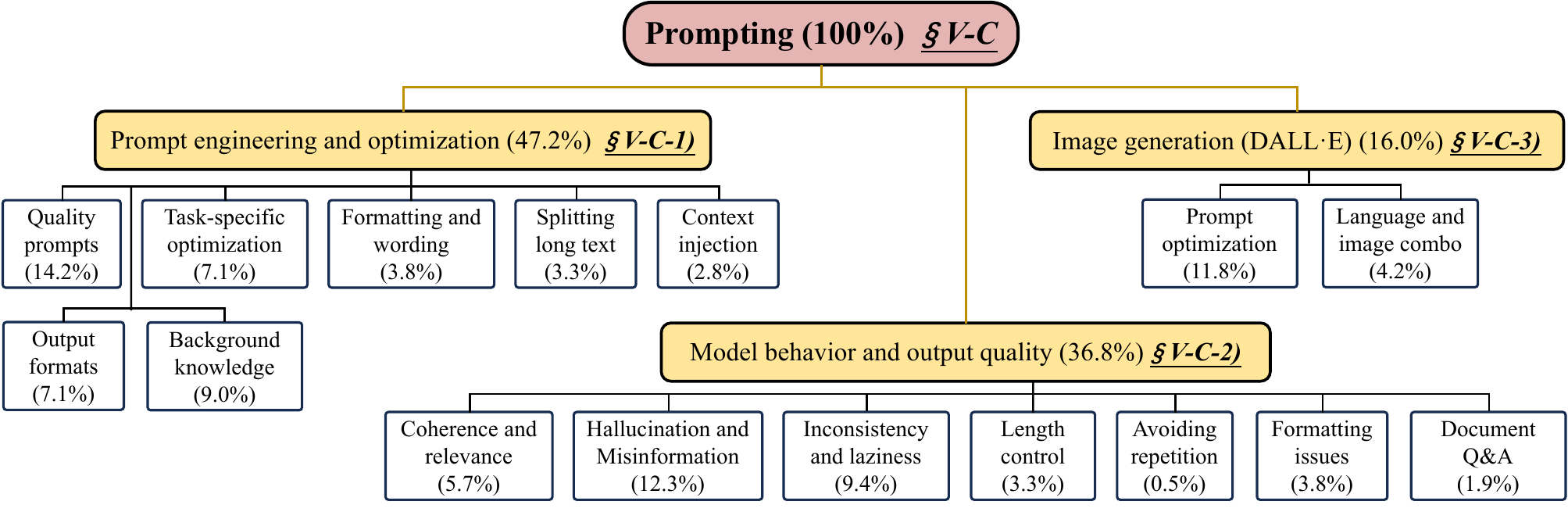}
        \caption{Taxonomy of concerns in \texttt{Prompting} topics.}
        \label{fig:prompting}
    \end{subfigure}
    
    \caption{Taxonomy of concerns on OpenAI Developer Forum.}
    \label{fig:taxonomy}
\end{figure*}

In this section, we construct a taxonomy based on 886 sample topics from the OpenAI Developer Forum, as shown in \autoref{fig:taxonomy}. 
The taxonomy comprises three main categories: \texttt{API/ChatGPT} (291 topics), \texttt{GPT builders} (214 topics), and \texttt{Prompting} (212 topics). Additionally, 169 topics were unrelated to developer concerns and excluded from further analysis.
The percentages in \autoref{fig:taxonomy} represent the proportion of topics in each subcategory relative to the total number of topics in that category.

\subsection{API / ChatGPT}
\label{sec:api}

As shown in \autoref{fig:api}, the \texttt{API} and \texttt{ChatGPT} categories encompass several key areas of interest. The most prevalent concern, accounting for nearly half of all discussions, is \textbf{API issues} (49.5\%). The second most significant area, \textbf{specific functionality} (26.5\%), focuses on specialized features and their practical applications. \textbf{Model performance} (18.5\%) is the third major topic, addressing the efficiency and accuracy of generated outputs. Lastly, \textbf{security, privacy, and bias} (5.5\%) cover critical aspects of data protection and ethical usage.

\noindent\textbf{\textit{1) API issues}}

Developers frequently encounter various API issues, including \textbf{request and response errors}, \textbf{rate limits and quotas}, \textbf{streaming responses}, and \textbf{file processing} challenges. For instance, the GPT 3.5-Turbo API can randomly hang indefinitely~\cite{293385}, and a bug in the \texttt{Assistant API} causes the temperature to default to 1 when set to 0, while other values work as intended. Errors such as ``Error: 413 The data value transmitted exceeds the capacity limit'' when calling v1/images/edits~\cite{464513} and inappropriate function calls from the GPT-3.5-turbo-1106 model~\cite{603102} are also reported. Rate limits and quotas cause substantial delays, with response times averaging around 30 seconds~\cite{547875}, and issues like insufficient quota errors on paid accounts~\cite{717097} and billing bugs~\cite{706641} persist. Streaming responses face performance decreases with large instructions~\cite{851660}, missing characters~\cite{797939}, and issues handling image URLs~\cite{748357}. File processing problems include the code interpreter's inability to find uploaded files~\cite{856370}, persistent file appearance in lists despite deletion~\cite{738145}, and errors like ``openai.BadRequestError: Error code: 400'' during uploads~\cite{769418}. 

\noindent\textbf{\textit{2) Model performance}}

Many posts report \textbf{a noticeable decline in GPT-4 and GPT-4o's performance, especially via the API}, over the past few months~\cite{829203}. Developers have experienced issues with context retention, code generation, and accuracy, similar to those in models like GPT-3.5-turbo and DALL-E. For example, GPT-4o often forgets its own modifications when generating Python code and fails to follow simple instructions~\cite{829203}. It also introduces inconsistent logic and references non-existent variables and functions~\cite{829203}. Additionally, there are complaints about outdated or incorrect code~\cite{805602}. Developers have also reported instances of \textit{\textbf{over-refusals, where GPT-4o and other models refuse to perform tasks well within their capabilities}}~\cite{409799}. Common examples include refusing to handle PDF files, generate images or charts, or run code due to assumptions about missing Python packages. For instance, GPT-4 Turbo refused to refactor code due to an assumed platform limit, even though the task was within token limits~\cite{548716}. Another example involved the model refusing to directly convert Lua scripts to Python, requiring multiple prompts to clarify the request~\cite{548716}. Developers have also noted instances where the model refuses to work with ``copyrighted material'' or denies having analytical tools, even when such tasks should be feasible, leading to significant productivity loss~\cite{671453}.

\find{\textbf{Insight 1:}
Increasingly, state-of-the-art LLMs~\cite{cui2024or,xie2024sorry} exhibit over-refusals, suggesting a potential research direction to explore the root causes of these occurrences and how to mitigate them.
}

\noindent\textbf{Platform discrepancy.}
Significant performance differences exist between the API and web-based (ChatGPT) versions of GPT-4. Despite identical prompts and parameters, the API output is often considered less intelligent or relevant compared to the web version. For example, the GPT-4 API's outputs are described as ``way worse'' than those from ChatGPT on the website, even after experimenting with system prompts and different temperatures~\cite{253218}. The API sometimes fails to retrieve information from files, returning responses like ``I don’t have any information about this''~\cite{722245}. Additionally, the GPT-4 Turbo model generates random garbage text when using the async API, an issue not seen with GPT-4o or GPT-3.5 Turbo~\cite{791641}. Concerns also exist about the API embedding engine performing worse than the one used by ChatGPT~\cite{780947}. \autoref{tab:issues_across_platforms} illustrates that various ChatGPT applications on different platforms have their own specific issues. For instance, audio responses on the iOS app can be choppy and unclear, making it difficult to understand the voice content~\cite{815847}. On the Mac OS app, developers cannot view conversation text when using the voice feature~\cite{792894}. Additionally, there are problems with image generation/upload on the Mac and iOS apps, and inconsistencies in accessing old responses on Android.

\find{\textbf{Insight 2:}
The API version of GPT-4 often underperforms compared to the web version (ChatGPT). Furthermore, there are noteworthy issues with ChatGPT applications across different platforms, as shown in \autoref{tab:issues_across_platforms}.
}

\input{Tables/chatgpt}

\noindent\textbf{\textit{3) Specific functionality}} 

Several widely-used functionalities, including \textbf{DALL-E}, \textbf{Whisper}, \textbf{TTS-1}, and \textbf{Embeddings}, have encountered various issues. There is a significant drop in quality when using the \texttt{DALL-E API}, particularly with DALL-E3, where results are noticeably inferior to those from Bing and the original model~\cite{492970}. The \texttt{Whisper API}'s performance has deteriorated over time, showing hallucinations and dropped sentences, especially with significant background noise. Attempts to clean audio files using \texttt{ffmpeg} have been insufficient, and developers suggest appending a pattern to the prompt to filter out random text generated during silent segments~\cite{473368}. The \texttt{TTS-1 API} struggles with text extraction from PDFs and produces tinny, distorted voices compared to the standard offering. Developers recommend first converting PDFs to text files and cleaning them up before processing through the \texttt{TTS} model. Additionally, there are issues with incomplete audio outputs and inconsistent responses for non-English requests~\cite{766983}. \texttt{Embeddings} are cost-efficient, but developers face batch processing rejections due to a full queue, inconsistent \texttt{embeddings} for the same content, and performance differences between small and large models. These issues impact tasks such as search, clustering, recommendations, anomaly detection, diversity measurement, and classification~\cite{742152}.

\noindent\textbf{Multimedia processing.} 
Developers have reported several problems when processing multimedia using OpenAI's API. One common issue is the ``invalid\_image'' error during batch processing of images stored in public S3 buckets, despite the images being valid and accessible~\cite{859709}. This issue appears random, with different requests failing in each batch. Some developers speculate that the API struggles with certain JPEG file structures and use services like Cloudinary to preprocess images before sending them to OpenAI's API~\cite{849277}. Additionally, intermittent failures occur when generating images with the \texttt{Assistant API} using a code interpreter to create graphs from CSV data, even though the code executes correctly, suggesting potential provisioning issues with the sandbox environments~\cite{713062}.

\noindent\textbf{Text formatting.} 
The AI sometimes fails to emit its stop sequence in ``JSON-mode'', leading to excessive text generation, and the GPT-4o API often produces invalid JSON responses~\cite{749391}. Additionally, LaTeX formatting for matrices fails consistently~\cite{853837}, and copying text from Microsoft Word into the Mac desktop app results in the text being pasted as an image~\cite{854119}. Uploading ``.js'' files also causes \textit{BadRequestError} (error code 400)~\cite{769418}.

\noindent\textbf{Conversation history management.} 
Developers have reported issues with the \texttt{Assistant API} where \textit{\textbf{responses are influenced by instructions from previously deleted assistants}}~\cite{863634}. This problem manifests even after creating new threads and deleting old ones. In some cases, the \texttt{assistant}'s responses are skewed towards instructions from the earliest \texttt{assistant} created in the project. To mitigate this, developers have added extensive new instructions and used completely new \texttt{assistant} names and rephrased instructions for each iteration. Despite these efforts, the issue raises concerns about potential guardrails between different \texttt{assistants}.
\textit{\textbf{This issue reveals a disregard for user rights by OpenAI.}} The persistence of deleted instructions undermines user control and data privacy, raising serious ethical concerns about the platform's respect for user autonomy and consent. Such behavior not only violates basic principles of responsible AI development but also potentially infringes on data protection regulations, highlighting a critical need for transparency and accountability in OpenAI's data management practices.

\find{\textbf{Insight 3:}
The SE community can focus on developing robust AI integration frameworks and advanced testing methodologies for AI-powered applications. These efforts may address cross-platform inconsistencies, enhance multimedia processing, and refine conversation management.
}

\noindent\textbf{\textit{4) Security, privacy, and bias}}

\noindent\textbf{API abuse and attacks.} 
Developers have reported \textit{\textbf{significant security vulnerabilities that have persisted for over a year}}, allowing bad actors to hack accounts, create organizations using the victim's payment method, and demote the original developer to ``Reader'' status~\cite{299460}. This demotion means the attacker takes over the ``Owner'' role within the organization, gaining full control over account settings, billing, and access management, while the ``Reader'' cannot affect the ``Owner's'' actions. 
This vulnerability may be caused by the developer's key leakage and flaws in OpenAI's authentication mechanism. 
Unlike OpenAI, other cloud platforms such as AWS, GCP, and Microsoft Azure implement stricter security configurations, including mandatory MFA for sensitive operations. It is crucial for developers to \textit{\textbf{adopt best practices in key distribution}} and for OpenAI to
enhance security measures to prevent such abuses and protect developer accounts.

\find{\textbf{Insight 4:}
Persistent security vulnerabilities in the OpenAI API account highlight the need for stricter security measures and best practices to prevent account takeovers and protect developer accounts.
}

\noindent\textbf{Biased content generation.}
Concerns have been raised about bias in OpenAI's products like DALL-E and ChatGPT, particularly regarding restrictions on generating illustrations for ``Eid al-Adha'' with Islamic symbols~\cite{822373}. While the system cites content restrictions to avoid misuse, similar symbols from other religions (e.g., Christmas, Easter, Star of David) do not face the same limitations. This inconsistency suggests unequal treatment of cultural symbols, prompting calls for OpenAI to review and address these biases to ensure fairness.

\subsection{GPT builders}
\label{sec:gpts}

The taxonomy of concerns within the \texttt{GPT builders} category encompasses several key areas, as shown in \autoref{fig:gpt_builders}.
\textbf{GPT builders}, accounting for 27.1\% of the concerns, focus on creating and managing various GPTs. \textbf{Action builders} represent 23.4\% of the concerns, dealing with API interactions and security measures. \textbf{Performance and errors} constitute 20.6\%, addressing response speeds and error handling. \textbf{GPT management} accounts for 15.9\%, involving the discovery and sharing of GPTs as well as handling feature requests. \textbf{Integration and deployment} make up 7.9\%, covering the incorporation of GPTs into web and mobile applications. Although representing 5.1\% of the concerns, \textbf{security and privacy} remain crucial, focusing on intellectual property (IP) protection and regulatory compliance.

\noindent\textbf{\textit{1) GPT builders}}

\noindent\textbf{Instruction following.} 
Custom GPTs often struggle to follow instructions consistently. For example, despite being within the 128,000 token limit, some custom GPTs ignore certain instructions, causing frustration~\cite{529666}. There are instances where GPTs disregard instructions regularly, affecting their reliability~\cite{551858}. To mitigate these issues, it is recommended to provide concise and clear instructions, use specific dialog starters, and structure the instructions effectively.

\noindent\textbf{Knowledge base.} 
Issues with the knowledge base functionality are common among GPT builders. Files uploaded to the GPT's knowledge section occasionally disappear or are not utilized during interactions~\cite{546431}. In some cases, financial GPTs fail to use the provided Quickbooks API knowledge, leading to incorrect query generation~\cite{569057}. Integrating external databases via REST API can be technically challenging, although there are solutions to simplify this process~\cite{580313}. Additionally, there are discrepancies between preview and live performance, with GPTs struggling to access and use data files consistently~\cite{552586}. These challenges highlight the need for improved reliability and integration of knowledge bases in custom GPTs.

\find{\textbf{Insight 5:}
A potential research direction could explore performance improvements of GPTs with integrated knowledge bases compared to ChatGPT and identify failure points when the knowledge base is not effectively utilized.
}

\noindent\textbf{\textit{2) Action builders}}

GPT \texttt{Actions} allow developers to interact with third-party services by executing API calls, converting input to JSON schemas, and handling specified authentication.

\noindent\textbf{API calls and responses.}
Issues with API calls and responses are common in GPT development. Problems include unsupported content types, inconsistent success between tools like Postman and GPT, and hallucinations of parameter values. Errors such as \textit{UnrecognizedKwargsError} or \textit{ApiSyntaxError} occur when GPT-generated requests deviate from the expected format~\cite{169600}. Additionally, the requirement for matching root domains in OAuth2 setups can hinder integration with services like Yahoo~\cite{491072}. These challenges necessitate robust debugging tools and more flexible API handling mechanisms. 

\noindent\textbf{Authentication and authorization.} 
Developers frequently encounter significant challenges related to authentication and authorization when integrating GPT with third-party services. Issues such as changing callback URL, infinite auth loops, and OAuth errors like “Missing token”~\cite{601033,493236} are common. Problems with API key validation, misconfigured endpoints, and permissions-policy headers further complicate the process~\cite{604653,531822}. 

\find{\textbf{Insight 6:}
For developers with limited programming experience, these obstacles make seamless integration challenging, suggesting the need for a comprehensive support toolset to facilitate smoother API calls and authentication handling.
}

\noindent\textbf{File processing errors.}
Common issues include validation errors indicating improperly passed parameters, multipart/form-data upload failures leading to 422 errors despite successful tests with tools like cURL and Postman, and empty responses when trying to download files through redirects~\cite{848948}. Additionally, there are problems with GPT not recognizing uploaded files in the conversation context, and generated download links being non-clickable in responses~\cite{641616}. These difficulties highlight the need for improved documentation and tools to streamline file-handling processes in GPT integrations.

\noindent\textbf{\textit{3) Performance and errors}}

Developers frequently encounter \textbf{delays in GPT response times}, particularly when using services like Ngrok, which can introduce over 30 seconds of latency~\cite{857902}. Additionally, there is a need for features to track the percentage of received responses during streaming to enhance user experience by buffering appropriately~\cite{775167}. \textbf{Error handling} is also challenging in GPT integrations, especially when dealing with large responses that exceed processing limits~\cite{500506}. For instance, custom \texttt{Actions} like accessing Google Calendar events can result in responses too large for GPT to handle, regardless of time range limitations~\cite{500506}. Furthermore, \textbf{token limits} present significant hurdles for GPT development. Developers face frustrations with request caps, which can halt progress for hours after sending a relatively small number of messages~\cite{575136}. The lack of guidance on managing these limits effectively leads to widespread developer dissatisfaction.

\noindent\textbf{\textit{4) Integration and deployment}}

Developers have reported several critical issues affecting the integration and deployment of GPTs with \texttt{Actions}. URLs served by custom \texttt{Actions} are not clickable on desktop but work fine on mobile devices, with errors related to the Permissions-Policy header~\cite{574476}. Additionally, a critical bug has affected GPTs with \texttt{Actions} on the ChatGPT app for Android since early May, causing them to malfunction despite successful API interactions, which does not occur in the browser version~\cite{737472}. There are also issues with function calling, particularly on Android devices like Samsung, where permissions and security settings cause persistent errors~\cite{857325}. 

\find{\textbf{Insight 7:}
Integration of GPTs with \texttt{Actions} face critical issues, especially on Android, and accessing custom GPTs via APIs or external applications is a possible extension.
}

\noindent\textbf{\textit{5) GPT management}}

\textbf{The discoverability of GPTs} in the store is currently limited, as only the top 12 GPTs are displayed, making it difficult for users to find lesser-known but valuable GPTs. Suggestions include showing all GPTs in a category with filtering options and implementing user recommendations based on usage history~\cite{585652}. Additionally, promoting GPTs with added value, such as those featuring \texttt{Actions} and knowledge files, over those popular due to their names, could help users explore and benefit from enhanced functionalities~\cite{585652}. Furthermore, developers have reported issues with \textbf{sharing custom GPT} links, including login loops and redirection to the basic ChatGPT 3.5 interface instead of the intended custom GPT, causing confusion and undermining the value of custom GPTs, especially for non-Plus users~\cite{502216}. There are also \textbf{feature requests} for enhanced capabilities, such as allowing AI to interact with web content more effectively by segmenting web pages with accessibility grids and providing dashboards to show statistics of their GPTs directly from the GPT website or app~\cite{755741}.

\find{\textbf{Insight 8:}
Exploring methods to enhance the added value of GPTs and optimizing recommendation systems are crucial for improving discoverability and user engagement.
}

\noindent\textbf{\textit{6) Security and privacy}}
\label{sec:gptsecurity}

\noindent\textbf{IP protection.}
Protecting the intellectual property (IP) of GPT models is a significant concern for developers. \textit{\textbf{One major risk is the potential for instructions to be stolen, allowing others to create and publish similar GPTs.}} Developers are particularly concerned that their carefully crafted instructions could be exposed if system prompt protections are bypassed~\cite{770720}. Issues such as instruction leakage and prompt injection can compromise the security and privacy of custom GPTs~\cite{616927}. Despite robust strategies, it is difficult to ensure complete immunity against ``cracking'' attempts. Developers must stay vigilant and continuously update their protection methods. Community contributions and collaborative efforts are vital in refining these security measures~\cite{587766}.\textit{\textbf{ Implementing structured prompts and utilizing tools like Prompt-Defender can provide additional layers of security}}~\cite{725650}.

\find{\textbf{Insight 9:}
Protecting GPT intellectual property is crucial, requiring constant vigilance and updates against risks like instruction theft and prompt injection.
}

\noindent\textbf{Data breach.}
Security and privacy breaches between plugins present a significant risk. For instance, users may inadvertently share sensitive data across plugins, such as confidential medical information between DrSmithPlugin and DrJohnsonPlugin, or legal strategies between HusbandLawyerPlugin and WifeLawyerPlugin~\cite{469532}. This innate issue arises from either granting ChatGPT the autonomy to route messages for optimal performance or restricting it to protect user privacy. Although plugins have been discontinued, similar concerns may arise with GPTs. 
If a user engages multiple GPTs within a single conversation, there is a risk of data being shared between them. Additionally, enabling multiple third-party services within GPTs may lead to data being shared across these services. 

\find{\textbf{Insight 10:}
Data breaches between GPTs and third-party services may pose risks, necessitating stringent privacy protections to address potential data-sharing concerns.
}

\noindent\textbf{Compliance and regulations.}
Navigating compliance and regulatory issues is crucial for GPT development and deployment. Developers face challenges such as unexpected GPT removals due to policy violations, prolonged appeal processes, and functionality restrictions possibly influenced by competitive strategies~\cite{740289}.

\subsection{Prompting}
\label{sec:prompting}

The concerns within the \texttt{Prompting} category, as shown in \autoref{fig:prompting}, encompass three primary areas. \textbf{Prompt engineering and optimization}, accounting for 47.2\% of the concerns, focuses on enhancing prompt quality, task-specific optimization, and appropriate formatting. \textbf{Model behavior and output quality} represents 36.8\% of the concerns, addressing critical issues such as coherence, relevance, hallucination, and misinformation in model outputs. The remaining 16.0\% is dedicated to \textbf{image generation} (DALL-E), which involves prompt optimization specifically for image creation and the integration of language with visual elements.

\noindent\textbf{\textit{1) Prompt engineering and optimization}}

Many developers have expressed concerns about the quality of prompts used in LLMs. Creating high-quality prompts is essential for eliciting accurate and relevant responses. \textbf{Task-specific optimization} involves tailoring prompts to each task's unique requirements, ensuring the AI understands the context and desired outcome~\cite{861669}. Proper formatting and wording can significantly impact prompt clarity and effectiveness, often requiring iterative refinement~\cite{853059}. \textbf{For lengthy texts, splitting them into manageable sections} helps maintain coherence~\cite{854697}. \textbf{Context injection} involves embedding relevant background information directly into the prompt~\cite{852560}. \textbf{Specifying output formats} clearly ensures the generated content meets expected standards~\cite{866883}. Providing \textbf{background knowledge} within the prompt can enhance the AI's understanding and output quality~\cite{766021}. Developers face challenges with version control and testing prompt variations, often relying on direct API calls and GitHub for sharing~\cite{305309}. There's a demand for specialized tools for prompt management and testing, akin to a ``CodePen for prompts''~\cite{305309}. Techniques like chunking and context summarization help manage long content without exceeding token limits. 

\find{\textbf{Insight 11:}
There is a need for specialized tools and techniques to optimize prompts and manage prompt variations for better LLM responses.
}

\noindent\textbf{\textit{2) Model behavior and output quality}}

Developers have raised concerns about \textbf{coherence and relevance issues, hallucinations, inconsistency, and laziness in model outputs}. GPT-4 has been reported to struggle with rotated text in images~\cite{461827}, forget the previous context, and generate nonsensical or incomplete outputs despite negative feedback. Hallucinations have been observed when processing non-structured data from CSV files~\cite{22280} and when chat history becomes too long or there are too many input sources~\cite{244698}. Inconsistency in performance has been noted, with ChatGPT providing different answers to the same task or refusing to complete tasks after a certain number of requests~\cite{664193}. Models have shown reluctance to browse web pages, provided overly lengthy responses with unsolicited code, and omitted necessary example code when updating~\cite{688884}. Complex prompts have led to deteriorating results, and models have been found to override user-defined instructions, resulting in suboptimal outputs that fail to address users' needs comprehensively~\cite{742558}.

Another set of concerns raised by developers revolves around \textbf{length control, repetition avoidance, formatting issues, and document question-answering}. Users have reported difficulties in controlling the output length of models, with GPT-4 generating overly long responses or failing to extract all relevant categories from a given text. Repetition has been observed in model outputs, particularly when dealing with low-quality audio signals in transcription tasks~\cite{571739}. Formatting issues have arisen, with models including unwanted introductory phrases or failing to adhere to the desired output format, such as bullet points or tables~\cite{821841}. Lastly, developers have noted inconsistencies and inaccuracies in document question-answering tasks, with models fabricating information, overlooking specific details, or struggling to process large JSON files effectively~\cite{524103}.

\find{\textbf{Insight 12:}
Robust context management, advanced prompt engineering, and innovative approaches for dynamic output adjustment based on user feedback are needed. The community should explore efficient architectures for diverse data processing and develop comprehensive evaluation frameworks to improve model performance across tasks.
}

\noindent\textbf{\textit{3) Image generation}}

\noindent\textbf{Prompt optimization.}
Developers have expressed frustration with DALL-E's inability to consistently follow explicit instructions, such as avoiding the addition of text in generated images~\cite{570654}, maintaining consistency in character appearance and key objects across a sequence of images~\cite{691466}, or adhering to specific lighting and shadow requirements~\cite{829022}. Users have also reported difficulties in generating full-body images of characters despite providing detailed descriptions~\cite{786849}.

\noindent\textbf{Language and image combo.}
Challenges have arisen in seamlessly combining text generation and image generation within a single prompt, with developers struggling to create a coherent flow between the two modalities~\cite{693200}. Additionally, users have encountered issues with DALL-E when attempting to create infographics, noting that the generated statistical data and text often contain errors, omissions, and spelling mistakes, requiring significant manual editing to rectify~\cite{507925}. Developers have also sought guidance on extracting data from graphical images using ChatGPT.

\find{\textbf{Insight 13:}
Researchers should investigate prompt optimization differences between image and text generation, enhancing DALL-E's instruction adherence. Focus areas include improving image generation model interpretation, developing multi-modal integration frameworks, and exploring techniques for text-image coherence and accurate visual information representation.
}

%% file: Tables/chatgpt.tex
\begin{table}[h!]
\centering
\caption{Issues across different ChatGPT applications on various platforms.}
\resizebox{\linewidth}{!}{
\begin{tabular}{llccccc}
\toprule[1.2pt]
\textbf{Category} & \textbf{Issue} & \textbf{Win} & \textbf{Mac} & \textbf{Andr.} & \textbf{iOS} & \textbf{Web} \\ 
\midrule
Account \&  & Subscription/billing &  &  & \checkmark & \checkmark & \checkmark \\ 
Subscription & Login/authentication & \checkmark & \checkmark & \checkmark & \checkmark & \checkmark \\ 
& Limited GPT-4 access & \checkmark &  & \checkmark & \checkmark & \checkmark \\ 
& Access denied &  & \checkmark &  &  & \checkmark \\ 
\midrule
Performance \&  & Slow response &  & \checkmark & \checkmark &  & \checkmark \\ 
Technical & Errors/crashes & \checkmark & \checkmark & \checkmark & \checkmark & \checkmark \\ 
\midrule
Functionality & Image generation/upload &  & \checkmark &  & \checkmark & \checkmark \\ 
& Custom GPT/plugins & \checkmark & \checkmark & \checkmark & \checkmark & \checkmark \\ 
& PDF/file reading &  &  &  &  & \checkmark \\ 
& Copy/paste &  & \checkmark & \checkmark &  &  \\ 
& Search &  & \checkmark & \checkmark &  &  \\ 
& Model switching &  &  & \checkmark & \checkmark &  \\ 
& Voice chat & \checkmark & \checkmark & \checkmark & \checkmark &  \\ 
& Hyperlink retrieval &  & \checkmark &  &  &  \\ 
& Screen sharing &  & \checkmark &  &  &  \\ 
& Text-to-speech &  & \checkmark &  & \checkmark &  \\ 
& LaTeX rendering &  &  & \checkmark & \checkmark &  \\ 
& Data export/history &  &  & \checkmark &  &  \\ 
& Access old responses &  &  & \checkmark &  &  \\ 
& Voice widget &  &  & \checkmark &  &  \\ 
& Image loading &  &  &  & \checkmark &  \\ 
& Shortcuts integration &  &  &  & \checkmark &  \\ 
& Voice input time limit &  &  &  & \checkmark &  \\ 
& Specific voices &  &  &  & \checkmark &  \\ 
\midrule
UI/UX & UI/UX improvements & \checkmark & \checkmark &  & \checkmark & \checkmark \\ 
& Foreground, auto-scroll &  & \checkmark &  &  &  \\ 
& Desktop vs web differences & \checkmark & \checkmark &  &  & \checkmark \\ 
& Text area/font size & \checkmark & \checkmark &  &  &  \\ 
& Sidebar not showing &  &  &  &  & \checkmark \\ 
\midrule
Integration \&  & App/shortcut conflicts & \checkmark & \checkmark &  & \checkmark &  \\ 
Compatibility & Device compatibility &  & \checkmark  & \checkmark & \checkmark &  \\ 
& Android studio integration &  &  & \checkmark &  &  \\ 
\midrule
Privacy \&  & Content consistency &  &  &  &  & \checkmark \\ 
Data & Data retention &  &  &  &  & \checkmark \\
& App/extension data sharing &  &  &  & \checkmark &  \\ 
\midrule
Miscellaneous & Download availability & \checkmark & \checkmark & \checkmark & \checkmark &  \\ 
& Date/time settings &  & \checkmark &  &  &  \\ 
& Region/language support &  &  &  & \checkmark & \checkmark \\ 
\bottomrule[1.2pt]
\end{tabular}}
\label{tab:issues_across_platforms}
\end{table}

%% file: Sections/5.discussion.tex
\section{Implications}
\label{sec:discussion}

Our analysis of the OpenAI Developer Forum reveals implications for researchers, developers, and LLM providers. 

\subsection{Implications for Researchers}

The analysis reveals several promising research directions for the academic community. The phenomenon of over-refusals in state-of-the-art LLMs warrants investigation into its underlying causes and potential mitigation strategies. Developing robust AI integration frameworks and advanced testing methodologies for AI-powered applications is crucial for addressing cross-platform inconsistencies and improving multimedia processing and conversation management. 
Researchers could explore performance differences between LLMs with and without integrated knowledge bases, identifying failure points and optimization strategies. The field of prompt engineering offers opportunities for developing specialized tools to optimize prompts and manage variations, potentially improving LLM response quality and consistency.

\subsection{Implications for Developers}
Our analysis highlights key focus areas for developers. They should be aware of performance discrepancies between different versions of LLMs (e.g., API vs. web versions) and address cross-platform inconsistencies. Implementing security measures is critical for preventing account takeovers and safeguarding API credentials. Developers should adhere to industry best practices in key distribution and management. When working with custom GPTs and \texttt{Actions}, developers should be mindful of potential integration issues and explore ways to enhance GPT value and discoverability. Finally, developers should protect their intellectual property when creating custom GPTs, implementing safeguards against instruction theft and prompt injection attacks.

\subsection{Implications for LLM Providers}

We identify several areas for improvement for LLM providers (especially OpenAI). They should address over-refusals in their models, balancing caution with functionality. Consistency across model versions and platforms is crucial, with API versions performing comparably to web versions. Providers should offer comprehensive support toolsets and clear documentation to aid developers and improve GPT integration with \texttt{Actions} on mobile platforms. Privacy and security should be prioritized, implementing stringent protections for data-sharing and strengthening measures to protect developer accounts (e.g., mandatory MFA for sensitive operations). This also includes addressing critical issues like the persistence of deleted instructions in the \texttt{Assistant API}, which raises ethical concerns about user autonomy and data privacy. Providers must ensure proper isolation between different \texttt{assistants} and respect user rights in data management. Finally, providers should develop robust context management systems, advanced prompt engineering tools, and innovative approaches for dynamic output adjustment, along with comprehensive evaluation frameworks to enhance model performance across various tasks.

%% file: Sections/6.limitation.tex
\section{Threats to Validity}
\label{sec:limitation}

\noindent\textbf{Omission of topic selection.}
One limitation of our study is the potential omission of topics related to developer concerns during the topic selection process. We directly filtered out topics belonging to the \texttt{Announcements}, \texttt{Community}, \texttt{Forum feedback}, and \texttt{Documentation} categories based on the official description of topic categories. However, these categories may still contain some bug reports or other issues posted by developers. To avoid any impact on the completeness of our taxonomy, we manually examined these topics, including 1,126 belonging to \texttt{Community}, 27 to \texttt{Forum feedback}, and 102 to \texttt{Documentation}. 
Our examination revealed that the majority of developer concerns are \texttt{API/ChatGPT} related (69 topics), followed by \texttt{GPT builders} (36 topics) and then \texttt{Prompting} related concerns (18 topics).
Moreover, these concerns all fit within our taxonomy's categories, further validating its comprehensiveness.

\noindent\textbf{Subjectivity of researchers.}
Another limitation is the subjectivity introduced by the researchers during the manual analysis. To mitigate this threat, two inspectors independently analyzed and labeled the sample topics. Any conflicting results were discussed with an experienced arbitrator until a consensus was reached. 
Fortunately, the independent labeling process yielded a high inter-rater agreement, demonstrating the reliability of our coding schema and procedure.

%% file: Sections/8.conclusion.tex
\section{Conclusion}
\label{sec:conclusion}

In this paper, we comprehensively analyzed the OpenAI Developer Forum, addressing two main research questions: the popularity trends and a taxonomy of developer concerns. By quantitatively analyzing forum metadata and qualitatively categorizing developer discussions, we identified key areas of interest, engagement patterns, and specific concerns raised by developers. Our findings provide valuable insights for future research, tool development, and best practices, ultimately enhancing AI-assisted software engineering and promoting the responsible integration of AI technologies.